\documentstyle[sprocl]{article}

\def\simge{\mathrel{%
   \rlap{\raise 0.511ex \hbox{$>$}}{\lower 0.511ex \hbox{$\sim$}}}}
\def\simle{\mathrel{
   \rlap{\raise 0.511ex \hbox{$<$}}{\lower 0.511ex \hbox{$\sim$}}}}

\def\Journal#1#2#3#4{{#1} {\bf #2}, #3 (#4)}

\def\be{\begin{equation}}
\def\ee{\end{equation}}
\def\bea{\begin{eqnarray}}
\def\eea{\end{eqnarray}}
\def\ba{\begin{array}} 
\def\ea{\end{array}}   

\begin{document}

\title{\large \bf ABOUT $\ R\,$-\,PARITY \bigskip \\
AND \bigskip \\
THE \,SUPERSYMMETRIC \,STANDARD \,MODEL}

\author{\vskip .8truecm P. FAYET}

\address{\vskip .2truecm
Laboratoire de Physique Th\'eorique de l'Ecole Normale 
Sup\'erieure\,\footnote{
UMR 8549, Unit\'e Mixte du CNRS 
et de l'Ecole Normale Sup\'erieure.
\medskip \\
{\normalsize {\bf Preprint ~LPTENS - 99/31}}
}, 
\\ 
24 rue Lhomond,
75231 Paris Cedex 05, France,
\\E-mail: fayet@physique.ens.fr}

\maketitle\abstracts{\vskip 1.3truecm
We recall the obstacles which seemed, long ago, to prevent
one from viewing supersymmetry as a possible fundamental symmetry of Nature.
Is spontaneous supersymmetry breaking possible\,?
Where is the spin-$\frac{1}{2}\,$ Goldstone fermion of supersymmetry, 
if not a neutrino\,?
Which bosons and fermions could be related\,?
Can one define conserved baryon and lepton numbers in such theories, 
although they systematically involve self-conjugate Majorana fermions\,? 
If we have to postulate the existence of new bosons 
carrying $\,B\,$ and $\,L\,$ 
(the new spin-0 squarks and sleptons),
can we prevent them from mediating new unwanted interactions\,? 
\medskip\\
We then recall how we obtained the three basic ingredients 
of the Supersymmetric Standard Model:
1) the $\,SU(3)\times SU(2)\times U(1)\,$ gauge superfields;
2) the  chiral quark and lepton superfields; 
3) the two doublet Higgs superfields
responsible for the electroweak breaking, 
and the generation of quark and lepton masses.
\linebreak
The original continuous $``R$-invariance'' of this model
was soon abandoned in favor of its discrete version, $\,R$-parity,
so that the gravitino, and gluinos, can acquire masses
\,--\, gluinos getting their masses from supergravity, 
or radiative corrections.
\smallskip\smallskip\\
\hbox{$R$-parity} forbids unwanted squark and slepton exchanges,
and guarantees the stability of the ``lightest supersymmetric particle''.
It is present only since we restricted the initial superpotential 
to be an {\it \,even\,} function of quark and lepton superfields
(so as to allow for $B$ and $L$ conservation laws),
as made apparent by the formula
re-expressing $\,R$-parity as $\ (-1)^{2 S}\ (-1)^{(3B+L)}\,$.
\,Whether it turns out to be absolutely conserved, or not, 
$\,R$-parity plays an essential r\^ole in 
the phenomenology of supersymmetric theories, 
and the experimental searches for the new sparticles.
}

\newpage

\section{Introduction}
\label{sec:intro}

\vskip .1truecm

The algebraic structure of supersymmetry in four dimensions,
introduced in the beginning of the seventies by 
Gol'fand and Likhtman~\cite{gl},
Volkov and Akulov~\cite{va}, and Wess and Zumino~\cite{wz},
involves a spin-$\frac{1}{2}\,$ fermionic symmetry generator $\,Q$ 
satisfying the (anti)\,commutation relations:
\bea
\label{alg}
\cases{ \ \ \ba{cccc}
\{ \ \,Q , \ {\bar Q} \ \,\} \ \ &=& \
- \ 2\ \,\gamma_{\mu} \  P^{\mu} &, \vspace {0.3 true cm} \cr 
[ \,\ Q, \ P^{\mu} \ ] \ \ &=& 
  \ \ \ \ \ \ \ \ \ \ \ \ 0  \ \ \ \ \ &.
\ea
}\label{ss}           
\eea
This (Majorana) spin-$\frac{1}{2}\,$ supersymmetry generator $\,Q\,$ 
was originally introduced as relating fermionic with bosonic fields, 
in the framework of relativistic field theories.
It can potentially relate fermions and bosons in a physical theory, 
provided one succeeds in identifying physical fermions
and bosons that might be related under such a symmetry.
The presence of the generator of spacetime translations $\,P^\mu\,$ 
in the right-handside of the anticommutation relations (\ref{alg})
is at the origin of the relation of supersymmetry with general relativity 
and gravitation, since a locally supersymmetric theory 
must be invariant under local coordinate transformations.

\vskip .3truecm
The consideration of this algebraic structure, 
if it is to be taken 
seriously as a possible symmetry of the physics 
of fundamental particles and interactions, led us to
postulate the existence of new ``superpartners'' 
for all ordinary particles~\cite{ssm,ssm2}.
These still-hypothetical superpartners may be 
attributed a new quantum number called $\,R\,$-parity, 
which may be multiplicatively conserved in a natural way,
and is especially useful to guarantee the absence of 
unwanted interactions mediated by squark or slepton exchanges, 
that might otherwise be present.
As is well-known, the conservation (or non-conservation)
of $\,R$-parity is closely related with the conservation 
(or non-conservation) of baryon and lepton numbers, $\,B\,$ and $\,L\,$.
~A conserved $\,R$-parity also ensures the stability 
of the ``lightest supersymmetric particle'',
\,a good candidate to constitute a part of the Dark Matter 
that seems to be present in our Universe.

\vskip .3truecm

It may be worth to go back in time, and think a little bit more 
about supersymmetry, and the way it might be present in Nature.
The supersymmetry algebra (\ref{alg}) was introduced, 
in the years 1971-1973, by three different groups, 
with quite different motivations. 
Gol'fand and Likhtman~\cite{gl}, in their remarkable work published in 1971,
first introduced it in connection 
with parity-violation, presumably in view
of understanding parity-violation in weak interactions:
when the Majorana supersymmetry generator $\,Q_\alpha\,$ 
is written as a two-component 
chiral Dirac spinor (e.g. $Q_L\,$), one may have the impression that the 
supersymmetry algebra is intrinsically parity-violating 
(which, however, is not the case).
Volkov and Akulov~\cite{va} hoped to explain the masslessness of the neutrino 
from a possible interpretation as a spin-$\frac{1}{2}\,$
Goldstone particle, while Wess 
and Zumino~\cite{wz} wrote the algebra by extending to four dimensions the 
``supergauge'' (i.e. supersymmetry) transformations~\cite{2d}, 
and algebra~\cite{2dalg},
acting on the two-dimensional string worldsheet.

\vskip .3truecm

     The mathematical existence of an algebraic structure does 
not imply that it has to play a r\^ole as an invariance 
of the fundamental laws of Nature. 
Incidentally while supersymmetry is commonly referred to as
``relating fermions with bosons'', \,its algebra (\ref{alg}) does not even 
require the existence of fundamental bosons (and even less of the
superpartners that were introduced later)\,! \,With non-linear 
realizations of supersymmetry a fermionic field can be 
transformed into a {\it \,composite\,} bosonic field made of fermionic 
ones~\cite{va}. \,Still we shall concentrate on the framework 
of the ``linear realizations'' of the supersymmetry algebra, 
which seems to be the most promising one, so as to work within the 
context of renormalizable supersymmetric gauge theories
(as far as weak, electromagnetic and strong interactions are concerned).

\bigskip

\section{Nature does not seem to supersymmetric\,!}
\label{sec:na}

\vskip .1truecm

Could supersymmetry be present as a fundamental symmetry of
 Nature, or is it doomed to remain, as it initially seemed 
for some time,
as a mathematical algebraic structure, only\,? 
Can we use this symmetry to relate directly 
known bosons and fermions\,?
And, if not, why\,?
\,If known bosons and fermions cannot be directly related by supersymmetry, 
do we have to accept this as the sign that supersymmetry is {\it \,not\,}
a symmetry of the fundamental laws of Nature, and 
drop this apparently unsuccessful idea\,?
\,Indeed, in the early days of supersymmetry (around 1974 or so),
many obstacles seemed to make its physical realization impossible.

\vskip .3truecm

Some of the main obstacles are summarized by the following questions 
(\,Q\,i's\,) below.
The most obvious one comes from the fact that, while 
bosons and fermions should have equal masses 
in a supersymmetric theory (in the framework of the ``linear realizations''
of the supersymmetry algebra), this is obviously not the case in Nature.
Supersymmetry should then clearly be broken.
At first sight this does not necessarily seem to be a problem, 
since we are so used to deal with spontaneously broken symmetries, 
such as, in particular, spontaneously broken gauge theories.

\vskip .3truecm

But supersymmetry is a special symmetry, 
since the Hamiltonian \,-- which plays an essential r\^ole 
in the definition of a stable vacuum state with minimum energy --\,
appears in the right-handside of the anticommutation relations  (\ref{alg})
of the supersymmetry algebra. 
Actually this Hamiltonian can be expressed 
proportionally to the sum of the squares of the components 
of the supersymmetry generator, 
$\,H\,=\,\frac{1}{4}\ \sum_\alpha\,Q_\alpha^{\ 2}\,$.
~This shows that a supersymmetry-preserving vacuum state must 
have vanishing energy~\cite{iz}, while a state which is not invariant 
under supersymmetry could na\"{\i}vely be expected to have a larger, 
positive, energy.
Indeed, such a supersymmetry-breaking state actually corresponds, 
in global supersymmetry,
to a positive energy density, the scalar potential, written 
proportionally to the sum of the squares of the auxiliary 
$\,D, \ F\,$ and $\,G\,$ components as
$\ \,V\,=\,\frac{1}{2}\ \sum\ (\,D^2\,+\,F^2\,+\,G^2\,)\,$,
~being strictly positive.
As a result, potential candidates for
supersymmetry-breaking vacuum states
seemed to be necessarily unstable, 
with some of the spin-0 particles having negative mass$^2$,
~which is evidently not acceptable\,!
\,This initially seemed to make spontaneous supersymmetry-breaking impossible, 
leading to the question:
\bea
\hbox{Q}1:\ \ \ \ \ 
\hbox {\framebox [10cm]{\rule[-.3cm]{0cm}{.8cm} $ \displaystyle {
\hbox {Is spontaneous supersymmetry-breaking possible at all\,?}
}$}}
\nonumber
\eea

\vskip .3truecm
As we know several ways of breaking spontaneously 
global or local supersymmetry
have been found~\cite{fi,F,crem}.
\,But spontaneous supersymmetry-breaking remains, in general,
rather difficult to obtain, since theories tend to prefer, 
for energy reasons, supersymmetric vacuum states. 
Then, how can spontaneous supersymmetry-breaking be possible\,?

\vskip .3truecm
As we said above in global supersymmetry a non-supersymme\-tric state has, 
in principle, 
always more energy than a supersymmetric one, 
whose energy vanishes identically; 
it then seems that it should be unstable,
the stable vacuum state being, necessarily, a supersymmetric one\,!
\,Still it is possible to escape this general result 
\,-- and this is the key to 
supersymmetry-breaking --\,
if one can arrange to be in one of those rare situations 
for which {\it \,no supersymmetric 
state\,} {\it exists at all\,}
\,-- the set of equations 
$\ <\!D\!>'\hbox{s}\,=\,\,<\!F\!>'\hbox{s}\,=\,<\!G\!>'\hbox{s}\,=\,0\ $ 
having {\it \,no solution at all\,}.
\,But these situations are in general quite exceptional.
(This is in sharp contrast with ordinary gauge theories, for which one only
has to arrange for non-symmetric states to have less energy 
than symmetric ones, in order to get spontaneous symmetry-breaking.)

\vskip .3truecm
These rare situations usually involve an abelian $\,U(1)\,$ 
gauge group~\cite{fi}, 
allowing for a gauge-invariant linear \,``$\,\xi\,D\,$''\, 
term to be included in the Lagrangian density\footnote{Even 
in the presence of such a term, one does not necessarily get 
a spontaneous breaking of the supersymmetry: one has to be very careful 
so as to avoid the presence
of supersymmetry-restoring vacuum states, which generally tend to exist.
In a physical theory, whatever is the mechanism of 
supersymmetry breaking considered, 
one will have to check carefully, in particular, for the non-existence of
stable vacuum states for which electric charge and/or color symmetry 
would be spontaneously broken.
};
\,and/or an appropriate set of chiral superfields with very
carefully chosen superpotential interactions (``$F$-breaking'')~\cite{F}.
\,In local supersymmetry~\cite{sugra}, which includes gravity,
one has to arrange, at the price of a very severe fine-tuning,
for the energy density of the vacuum to vanish exactly~\cite{crem}, 
or almost exactly, to an extremely good accuracy,
so as not to generate an unacceptably large value of the 
cosmological constant $\,\Lambda\,$.

\vskip .3truecm

Once we know that it is possible to break supersymmetry spontaneously,
we shall still have to break it
in an acceptable way, so as to get  \,-- if this is indeed possible --\,
a physical world which looks like the one we know\,!
(The above $\,U(1)\,$~\cite{fi}, in particular, can{\it not\,} be identified 
with the weak hypercharge $\,U(1)\,$ of a physically-meaningful 
theory~\cite{extrau}, but might have been, 
instead, a new ``extra $\,U(1)$'' gauge symmetry~\cite{ssm}.)
\,Of course just accepting explicit supersymmetry-breaking 
without worrying too much about the origin of supersymmetry-breaking terms
would make things much easier
\,-- also at the price of introducing a large number of arbitrary parameters, 
coefficients of these supersymmetry-breaking terms.
But such terms must have their origin 
in a spontaneous supersymmetry-breaking mechanism, 
if we want supersymmetry to play a fundamental role, 
especially if it is to be realized as a local fermionic gauge symmetry, 
as in the framework of supergravity theories.

\vskip .5truecm
Another question to be asked immediately after learning
that spontaneous supersymmetry breaking is indeed possible, is:
\bea
\nonumber
\hbox{Q}2: \ \ 
\hbox {\framebox [10cm]{\rule[-.3cm]{0cm}{.8cm} $ \displaystyle {
\hbox {Where is the spin-$\frac{1}{2}\,$ 
Goldstone fermion of supersymmetry\,?}
}$}}
\eea

\noindent
The spontaneous breaking of the global supersymmetry must
generate a massless spin-$\frac{1}{2}\,$ Goldstone particle.
Could it be one of the known neutrinos~\cite{va}\,?
A first attempt at implementing the idea 
within a $\,SU(2) \times U(1)\,$ electroweak model of ``leptons''~\cite{toy}
(of the ``electron'' sector), discussed later in section
\ref{sec:toy}, quickly illustrated that this idea could not be 
pursued very far, for many reasons:
the existence of several neutrinos (two were known at that time), 
the fact that attributing a Goldstone r\^ole to one of them (and which one\,?) 
would break the lepton/quark universality, the fact that a Goldstone particle
has couplings of a very particular type which have to do
with the boson-fermion mass-splittings (or mass$^2$-splittings) 
within the multiplets of supersymmetry~\cite{toy,ssm2} 
(couplings from which one cannot recover those
of a neutrino), lepton-number conservation, 
low-energy theorems~\cite{bdf}, etc..

\vskip .3truecm

So, if the Goldstone fermion associated with the spontaneous breaking 
of the global supersymmetry is not one of the known neutrinos, 
where is it, and why hasn't it been observed\,?
One might suggest that it could be an (almost-decoupled)
right-handed neutrino $\,\nu_R$,
~but, again, the idea cannot be pursued very seriously, 
very much for the same reasons that we just indicated.

\vskip .3truecm

So, where is the spin-$\frac{1}{2}\,$ Goldstone fermion of supersymmetry\,?
Today we tend not to think at all about the question,
since: 1) the use of soft terms breaking {\it \,explicitly\,}
the supersymmetry makes the question irrelevant; \linebreak
2) 
since supersymmetry has to be realized locally anyway, 
within the framework of supergravity~\cite{sugra}, 
the massless \hbox{spin-$\frac{1}{2}\,$} Goldstone fermion (``goldstino'') 
should in any case be eliminated 
in favor of extra degrees of freedom for a massive 
spin-$\frac{3}{2}\,$ gravitino~\cite{ssm2,crem}.
\,So there is no goldstino, but a massive gravitino instead.

\vskip .3truecm
But the same question now gets transformed into: 
where is the gravitino, and why has no one
either seen a fundamental spin-$\frac{3}{2}\,$ particle\,?
To discuss it properly we need to know how bosons and fermions 
could be associated under supersymmetry (cf. the subsequent question Q3).
But we can anticipate that the interactions of the gravitino, 
being proportional to the Newton constant 
$\,G_N \simeq 10^{-38}\ \,\hbox{GeV}^{-2}$, 
~should be absolutely negligible in particle physics
experiments, so that this particle, even if it could be produced, 
would in any case remain undetected.
This should indeed be true, if the gravitino is heavy.

\vskip .3truecm
However we might be in a situation for which the gravitino is light, 
maybe even extremely light, so that it
would still interact very much like the massless Goldstone fermion 
of spontaneously-broken 
global supersymmetry, according to the ``equi\-valence theorem'' 
of supersymmetry~\cite{ssm2}.
In that case, the gravitino could well have non-negligible interactions,
relevant in particle physics, so that we 
should ask again the same question, but now for the $\,\pm\frac{1}{2}\,$
polarization states of the massive spin-$\frac{3}{2}\,$ gravitino.
The answer may be given later, after we get to 
the Supersymmetric Standard Model (cf. sections {\ref{sec:ssm}
and \ref{sec:rinv}): 
if $\,R$-parity is conserved, the 
$R$-odd gravitino should be produced in association with another
$\,R$-odd superpartner
\,--\, but it now seems that these superpartners should all be rather heavy.
\,(One may also consider the pair-production of very light gravitinos, 
but it is normally strongly suppressed at lower energies.)

\vskip .5truecm

Leaving for the moment this question of supersymmetry breaking,
the next \,-- and equally obvious --\, question to be asked after questions Q1
and Q2, 
\,is 
\bea
\nonumber
\hbox{Q}3: \ \ 
\hbox {\framebox [10.6cm]{\rule[-.3cm]{0cm}{.8cm} $ \displaystyle {
\hbox {Which bosons and fermions could be related by supersymmetry\,?}
}$}}
\eea
\vskip .1truecm

\noindent
There seems to be no positive answer to this question since the bosons
and fermions that we know do not seem to have much in common \,--
excepted maybe for the photon and the neutrino, which are both electrically 
neutral and massless (or almost massless).
We shall come back to this point in \hbox{section \ref{sec:toy}.}

\vskip .3truecm

Furthermore, from a more general point of view, 
the number of (known) fermionic degrees of freedom 
is significantly larger than for bosonic ones.
Actually we know today six quarks and six leptons, corresponding 
with their antiparticles to 90 fermionic degrees of freedom.
On the other hand the bosons that we know for sure to exist 
(ignoring the spin-2 graviton and the still-undiscovered Higgs boson)
are the color-octet of gluons, the photon, and the $\,W^\pm\,$ 
and $\,Z\,$ gauge bosons, which altogether 
correspond to 27 degrees of freedom, only.
In addition, these fermions and bosons have different gauge symmetry 
properties, e.g. the spin-$\frac{1}{2}\,$ quarks are (charged) 
color triplets while the spin-1 gluons form a (neutral) color octet (so that
attempting to relate directly gluons with quarks, for example,
would necessitate extended supersymmetry generators 
carrying both color and charge,
requiring very large representations involving higher-spin fields).

\vskip .3truecm

Another question, to which we shall return in sections \ref{sec:ssm} to
\ref{sec:rbl}, 
is the question of the definition of conserved baryon and lepton numbers. 
It once appeared as a serious difficulty 
in supersymmetric theories, especially since they systematically 
involve self-conjugate Majorana spinors:

\vskip-.3truecm
\bea
\nonumber
\hbox{Q}4: \ \ 
\hbox {\framebox [10cm]{\rule[-.5cm]{0cm}{1.2cm} $ \displaystyle {
\ba{c} \hbox {How could one define conserved
} \\
\hbox{baryon and lepton numbers, in a supersymmetric theory ?}
\ea
}$}}
\eea

\vskip-.04truecm
\noindent
Indeed these quantum numbers 
are known (for the moment) to be {\it \hbox{\,carried by\,} fermions only\,}
(the familiar spin-$\frac{1}{2}\,$ quarks and leptons), 
and {\it \,not\,} by bosons.
If we do insist on this property there is no real hope to be able to define 
such conserved baryonic and leptonic numbers\,!
(Of course nowadays we are so used to deal with spin-0 quarks and leptons, 
carrying baryon and lepton numbers almost by definition, 
that we can hardly imagine this could once have appeared as a problem.)

\vskip .3truecm

The solution to the problem went through the acceptance of the idea 
of introducing a large number of new bosons, 
also carrying baryon and lepton numbers,  
despite the fact that $\,B\,$ and $\,L\,$ were then viewed as 
intrinsically-fermionic numbers.
One had to accept the new idea of having
{\bf \,baryon and lepton numbers also carried by bosons}\,!

\vskip .3truecm

But if new spin-0 bosons carrying baryon or lepton numbers
are introduced (i.e. the new spin-0 quarks and leptons),
their direct (Yukawa) exchanges between ordinary 
spin-$\frac{1}{2}\,$ quarks and leptons, if allowed, 
could lead to an immediate disaster, preventing us from getting a theory 
of weak, electromagnetic and strong interactions mediated by spin-1 
gauge bosons (and not spin-0 particles), 
with conserved $\,B\,$ and $\,L\,$ quantum numbers\,!
\bea
\nonumber
\hbox{Q}5: \ \
\hbox {\framebox [9cm]{\rule[-0,6cm]{0cm}{1.4cm} $ \displaystyle {
\ba{c} \hbox {How can we avoid unwanted interactions} \\ 
\hbox{mediated by spin-0 squark and slepton exchanges\,?}
\ea
}$}}
\eea

\noindent
Fortunately, we can naturally avoid the existence of 
such unwanted interactions, thanks to $\,R$-parity, which 
guarantees that squarks and sleptons can{\it not\,} be
directly exchanged between ordinary quarks ans leptons, 
allowing for conserved baryon and lepton numbers 
in supersymmetric theories.

\bigskip

\section{$R$-invariance and electroweak breaking, 
from an attempt to relate the photon with the neutrino.}
\label{sec:toy}

\vskip .1truecm

Let us now return to an early attempt at relating 
{\it \,existing\,} bosons and fermions together.
Despite the general lack of similarities between known bosons and fermions, 
we might still try as an exercise to see 
how far one could go in attempting 
to relate the photon with one of the neutrinos (say ``$\nu_e$''),
\,in the framework of a spontaneously-broken supersymmetric theory~\cite{toy}.
At the same time, the $\,W^-\,$ boson could be related with 
a would-be ``electron''. 
This early model also showed how it was possible to define a conserved 
``leptonic'' number \,--\, called $R\,$.
~At that time the definition of a conserved quantum number 
carried by Dirac fermions posed a rather severe problem
in supersymmetric theories, since these theories make an extensive use 
of Majorana spinors, 
e.g. the spin-$\frac{1}{2}\,$ partners of the \hbox{spin-1} gauge bosons, 
now called ``gauginos''.
\,In particular the fermionic partner of the photon
\,-- to be called later the photino --\,
is precisely described by such a {\it \,self-conjugate\,} Majorana spinor. 

\vskip .3truecm

If we want to try to identify this companion of the photon
as being a ``neutrino'', we need to understand 
how it could carry a conserved quantum number 
that we could interpret as a ``lepton'' number.
We also need to be able to reconstruct charged massive Dirac spinors 
from originally massless components, having, furthermore, 
different electroweak gauge symmetry properties.

\vskip .3truecm
In the case of this toy $\,SU(2) \times U(1)\,$ model of ``leptons'',
the solution is obtained through the definition of

\vskip .3truecm
\begin{center}
{{\it a continuous $\,U(1)\,$ $\,R$-invariance,}}
\end{center}

\vskip .3truecm
\noindent
which made it possible to define such 
a conserved ``leptonic'' number.
It had the property that one unit of this \,-- additive --\,
quantum number (called $\,R\,$!)
was carried by the supersymmetry generator $\,Q_\alpha\,$.
The ``electron'' and ``neutrino'' candidates were indeed
described by  massive and massless Dirac spinors, 
each of them carrying one unit 
of the conserved quantum number $\,R\,$.
\,This continous $\,U(1)\,$ $\,R$-invariance~\cite{toy},
which also guaranteed the masslessness of the ``neutrino'',
acted chirally on the Grassmann coordinate $\,\theta\,$ 
which appears in the expression of the various 
(gauge and chiral) superfields.

\vskip .3truecm

In this first attempt \,-- which essentially became later a part 
of the Supersymmetric Standard Model  --\,
Higgs doublets responsible for the electroweak breaking
were related with ``leptonic'' doublets under supersymmetry. 
But in the resulting model (which also included a ``heavy electron''
carrying $\ -\,1$ unit of the additive $R\,$ quantum number)
\,one chiral component of the charged ``electron'' field 
transformed as the lower member of an $\,SU(2)\,$ triplet 
(and the other as the lower member of a doublet)\footnote{This woud-be 
``electron'' was obtained from a charged left-handed ``gaugino'' field
\,($\,\tilde W_L^{\,-}\,$), 
which acquired a mass by combining with a charged right-handed ``higgsino'' 
field. The ``neutrino'' was then described by the left-handed
gaugino field $\ \tilde \gamma_L\,$.},
\,and the ``neutrino'' was not directly coupled to the $\,Z\,$ boson.
As we know now, this is not acceptable
(nor is it for the leptons of the two other families, 
the one of the muon and the one of the $\,\tau^-$).
~Furthermore, if we insisted on such a scheme for the leptons 
of the electron sector,
what should we do with the other leptons of the muon and $\tau$ sectors, 
and with the quarks\,?

\vskip .3truecm

It was clear from the beginning that attempting 
to relate the photon with one of the neutrinos
could only be an exercise of limited validity,
but it had the merit of illustrating how one can break spontaneously a
$\,SU(2) \times U(1)$ gauge symmetry in a supersymmetric theory, 
through an electroweak breaking induced by 

\begin{center}
{\it a \underline{pair} \,of  chiral doublet Higgs superfields},
\end{center}

\vskip .3truecm
\noindent
now known as $\,H_1\,$ and $\,H_2$ (or $\,H\,$ and $\,\bar H\,$).
~In modern language, our previous would-be ``electron'' and ``heavy electron''
were in fact what we now call two winos, or charginos,
obtained through a mixing of charged gaugino and higgsino components.
The associated mass matrix simply reads, in a gaugino/higgsino basis,

\be
{\cal M}\ \ =\ \ \pmatrix{  (\,m_2\,=\,0\,) & 
\ \displaystyle{\frac{g\,v_2}{\sqrt 2}\,= \,m_{W}\sqrt{2}\,\sin \beta \ }\,
\cr \cr
\displaystyle{\ \frac{g\,v_1}{\sqrt 2}\,=\,m_{W}\sqrt{2}\,\cos \beta \ } 
&  \mu\,=\,0  \cr } \ \ ,
\ee

\vskip .2truecm
\noindent
in the absence of a direct higgsino mass originating from a
$\ \mu\ H_1 H_2\ $ mass term 
in the superpotential.
\,This $\,\mu\,$ term, which would have broken explicitly 
the continuous $\,U(1)\,$ $\,R$-invariance intended to be
associated with the ``lepton'' number conservation law, 
was already replaced by a $\ \lambda\ \,H_1 H_2\,N\ $ trilinear 
coupling involving an 
{\it \,extra neutral singlet chiral superfield\,} $\,N\,$
\footnote{A gaugino mass term 
parametrized, in the case of $\,SU(2)$, by $\,m_2$, 
~which would also have violated 
the continuous $\,U(1)\,$ $\,R$-invariance (in addition to the supersymmetry), 
does not appear, at tree level, in global supersymmetry.
But it could be generated by radiative corrections,
as soon as the continuous $\,R$-invariance is no longer present; 
it is also allowed in spontaneously-broken {\it \,locally\,}
supersymmetric theories, as we shall discuss in section \ref{sec:grav}.
}:
\be
\mu\ \,H_1 H_2\ \ \ \longmapsto \ \ \ \lambda\ \ H_1 H_2\,N\ \ .
\ee

\vskip .15truecm

Let us note in passing that using only {\it \,one\,} 
doublet Higgs superfield $\,H$, describing a single {\it \,chiral\ }
higgsino doublet, which would now be denoted as, e.g.
$
\left(\  \ba{c} 
\tilde h^0 \\ \tilde h^- 
\ea \right)_L,
$
~~would have led to ``one and a half'' charged Dirac fermion,
namely a charged Dirac ``gaugino'' ($\,\tilde W^-\,$)
mixed with a {\it \ chiral\,\,} charged Dirac ``higgsino'' 
($\,\tilde h^-_{\,L}\,$),
~leaving us with a massless charged chiral fermion. This would be, 
evidently, unacceptable (even before having to take into consideration 
the corresponding anomalies, that would then be present, already,
in the quantum theory of {\it \,electromagnetism\,}).

\vskip .3truecm
The whole construction showed that one could deal elegantly 
with spin-0 Higgs boson fields
(not a very popular ingredient at the time)
in the framework of spontaneously-broken supersymmetric theories.  
Quartic Higgs couplings are no longer completely arbitrary, 
but get fixed by the values of the gauge coupling constants
\,\hbox{\,-- $g\,$ and $\,g'$ --\,}\, through the following ``$D$-terms''
\,(i.e. 
\small $\ \frac{\vec D ^2}{2}\,+\,\frac{D'^2}{2}\,$
\normalsize
)\,
in the scalar potential given in ~\cite{toy} 
(with a different denomination for the two Higgs doublets,
such that
$\ \varphi'' \ \mapsto \ h_1,\ (\varphi')^c\ \mapsto
\ h_2,$ $\ \tan \delta = v'/v''\ \mapsto\ \tan \beta = v_2/v_1\, $):
\be
\hbox {\framebox [11.1cm]{\rule[-.9cm]{0cm}{2cm} $ \displaystyle {
\ba{ccl}
\ V_{\hbox{\small{Higgs}}} \ &=&\ \displaystyle{\frac{g^2}{8}\ \ 
(\,h_1^\dagger\ \vec \tau \ h_1\,+\,h_2^\dagger\ \vec \tau \ h_2\,)^2 \ + \ 
 \frac{g'^2}{8}\ \ 
(\,h_1^\dagger\,h_1-h_2^\dagger\,h_2\,)^2 \ +\,...\ \ } \vspace{3mm}\\
&=& \ \
 \displaystyle{ \frac{g^2\,+\,g'^2}{8}\ \ 
(\,h_1^\dagger\,h_1\,-\,h_2^\dagger\,h_2\,)^2 \ \ +\ \
\frac{g^2}{2}\ \ |\,h_1^\dagger\,h_2\,|^2\ \ +\ \ ...\ \ . }
\ea
}$}}
\ee

\noindent
This is precisely the quartic Higgs
potential of the ``minimal'' version of the Supersymmetric Standard Model, 
the so-called MSSM, with its quartic Higgs coupling constants equal to
\be
\frac{g^2\,+\,g'^2}{8}\ \ \ \ \hbox{and}\ \ \ \ \ \frac{g^2}{2}\ \ .
\ee

\noindent
Further contributions to this quartic Higgs potential also appear
in the presence of additional superfields, such as the
neutral singlet chiral superfield $\,N\,$ already introduced in this model,
which will play an important r\^ole in the NMSSM, 
i.e. in ``next-to-minimal'' or ``non-minimal'' versions of 
the Supersymmetric Standard Model.

\vskip .4truecm

{\bf Charged Higgs bosons} (now called $\,H^\pm$) are present 
in this framework, as well as {\it \,several neutral ones\,}.
All this is at the origin of various mass relations
(equalities or inequalities) connecting Higgs masses
to gauge boson masses 
in supersymmetric theories. Their
particular expressions depend on the details 
of the supersymmetry-breaking mechanism considered:
soft-breaking terms, possibly ``derived from supergravity'', 
presence or absence of extra-$U(1)\,$ gauge fields 
and/or additional chiral superfields, use of radiative corrections, etc..

\bigskip

\section{From the electroweak breaking 
\,to the Supersymmetric Standard Model.}
\label{sec:ssm}

\vskip .1truecm
These two Higgs doublets $\,H_1\,$ and $\,H_2\,$ 
are precisely the two doublets 
which I used in 1977 to generate the masses of charged-leptons
and down-quarks, and of up-quarks, in supersymmetric 
extensions of the standard model~\cite{ssm}. 
~Note that at the time 
having to introduce Higgs fields was generally 
considered as rather unpleasant, at least.
While one Higgs doublet was taken as probably unavoidable 
to get to the standard model
or in any case simulate the effects of the spontaneous breaking of the
electroweak symmetry, having to consider two Higgs doublets, 
thereby necessitating charged Higgs bosons as well as several neutral ones,
was usually considered as a too heavy price, in addition to the 
``doubling of the number of particles'', once considered as
an indication of the irrelevance of supersymmetry.
Actually many physicists spent a lot of time, 
later on, trying to avoid fundamental \hbox{spin-0} Higgs fields and particles,
before returning to fundamental Higgses, 
precisely in this framework of supersymmetry.

\vskip .4truecm
In the previous $\,SU(2)\times U(1)\,$ model~\cite{toy}, \,it was clearly
impossible to view seriously for very long ``gaugino'' and ``higgsino'' fields 
as possible building blocks for our familiar lepton fields.
This becomes even more patent if one takes again
quarks and gluons into consideration.
This led us to consider that all quarks, and leptons as well,
should be associated with new bosonic partners, the 
{\bf \,spin-0 quarks and leptons}.
Gauginos and higgsinos, mixed together 
by the spontaneous breaking of the electroweak symmetry,
correspond to a new class of fermions, now known as ``charginos''
and ``neutralinos''.

\vskip .5truecm

In particular,
the partner of the photon under supersymmetry, which 
cannot be identified with any of the known neutrinos, 
should be viewed as a new ``photonic neutrino''
which I called in 1977 the {\bf \,photino\,};
the fermionic partner of the gluon octet is an octet of self-conjugate 
Majorana fermions called {\bf \,gluinos\,}, 
etc. -- although at the time {\it \,colored fermions\,} belonging to 
{\it \,octet\,} 
representations of the color $\,SU(3)\,$ gauge group were generally believed 
not to exist\,\footnote{One could even think of using the absence of 
such particles as a general constraint to select admissible 
grand-unified theories~\cite{gm}.}!

\vskip .3truecm
The two doublet Higgs superfields
$\,H_1$ and $\,H_2\,$
introduced previously are precisely those needed to generate 
quark and lepton masses in supersymmetric extensions 
of the standard model~\cite{ssm} \footnote{
The correspondance between our earlier notations 
for doublet Higgs superfields and mixing angle, 
and modern ones, is as follows:
\\
\begin{center}
\begin{tabular}{|ccc|} \hline 
&&\\ 
$S\ =\ \left( \ba{cc} S^0 \vspace{.1truecm}\\ S^-
\ea \right)\ \,\hbox{and}\ \ \,
T\ = \ \left( \ba{cc} T^0 \vspace{.1truecm}\\ T^-
\ea \right)$	 &  $\longmapsto $   &  
$H_1\ =\ \left( \ba{cc} H_1^{\,0} \vspace{.1truecm}\\ H_1^{\,-}
\ea \right)\ \,\hbox{and}\ \ \,
H_2\ = \ \left( \ba{cc} H_2^{\,+} \vspace{.1truecm}\\ H_2^{\,0}
\ea \right)$
                           \\  [.2 true cm] && \\
(left-handed) \ \ \ \ \ \ \ \ (right-handed) \  &   &   (both left-handed)   
         \\   [.2 true cm]   \hline  && \\ 
$ \hbox{\large{$\tan \,\delta$}} \,=\ \frac{<T^0>}{<S^0>}\ =\ 
\frac{<\varphi'^0>}{<\varphi''^0>}\ =\ \displaystyle{v'\over v''}$ & 
$ \longmapsto $ &
$ \hbox{\large{$\tan \,\beta$}} \,=\ \frac{<H_2^{\,0}>}{<H_1^{\,0}>}\ = \
\frac{<h_2^{\,0}>}{<h_1^{\,0}>}\ = \ \displaystyle{v_2\over v_1}
$
 \\ [.2 true cm] && \\ \hline 
\end{tabular}
\end{center}
},
in the now-usual way, through the familiar trilinear superpotential
\be
\label{supot}
{\cal W}  \ \ = \ \ h_e \ H_1 \,.\,\bar E \,L \ +\ 
h_d \ H_1\,. \,\bar D \,Q \ -\  
h_u \  H_2 \,.\,\bar U \,Q \ \ \ .			    
\ee
Here $\,L\,$ and $\,Q\,$ denote the left-handed doublet lepton and quark 
superfields, and $\,\bar E$, $\bar D\,$ and $\,\bar U\,$ left-handed singlet 
antilepton and antiquark superfields.
(We originally denoted, generically, by $\,S_i\,$, left-handed, 
and $\,T_j\,$, right-handed,
the chiral superfields describing the left-handed and right-handed 
spin-$\frac{1}{2}$ quark and lepton fields, 
together with their spin-0 partners.)
~The vacuum expectation values of the two Higgs doublets 
described by
$\,H_1\,$ and $\,H_2\,$ generate charged-lepton and down-quark masses, 
and up-quark masses,
given by
$\,m_e\,=\,h_e\,v_1/2\,,\ \,m_d\,=\,h_d\,v_1/2\,,$ ~and 
$\,m_u\,=\,h_u\,v_2/2\,$,
~respectively.

\vskip .3truecm

This constitutes the basic structure of the 
{\bf \,Supersymmetric Standard Model\,},
which involves, at least, the basic ingredients 
shown in Table \ref{tab:basic}.
Other ingredients, such as a direct $\,\mu\ H_1 H_2\,$ direct mass term 
in the superpotential, or an extra singlet chiral superfield $\,N\,$ 
with a trilinear superpotential coupling $\ \lambda \ H_1 H_2\,N\,+ \, ... \ $
possibly acting as a replacement for a direct $\,\mu\ H_1H_2\,$ 
mass term, as in ~\cite{toy},
~and/or extra $\,U(1)\,$ factors in the gauge group, 
may or may not be present, depending on the
particular version of the Supersymmetric Standard Model considered.

\begin{table}[t]
\caption{\ The basic ingredients of the Supersymmetric Standard Model.
\label{tab:basic}}
\vspace{0.2cm}
\begin{center}
\begin{tabular}{|l|} \hline \\ 
\ 1) the  three $\,SU(3)\times SU(2)\times U(1)\,$ gauge superfield
representations;  \ \\ \\
\ 2) the  chiral quark and lepton superfields corresponding  \\
\hskip 2truecm to the three quark and lepton families; \\ \\
\ 3) the two doublet Higgs superfields $\,H_1\,$ and $\,H_2\,$
responsible  \\ 
\hskip 2truecm for the spontaneous electroweak symmetry breaking,  \\
\hskip 2truecm 
and the generation of quark and lepton masses  \\
\hskip 3truecm through the trilinear superpotential (\ref{supot}).
\\  \\ \hline
\end{tabular}
\end{center}
\end{table}

\vskip .3truecm

It is often useful to know, in addition, that the gauge interactions
of the quark, lepton and Higgs superfields, 
and the trilinear superpotential interactions (\ref{supot})
responsible for quark and charged-lepton masses 
are also invariant under an {\it \,extra $\,U(1)\,$
symmetry\,}\,\footnote{This is precisely the extra $\,U(1)\,$ symmetry 
which we initially proposed to gauge~\cite{ssm}, 
in addition to $\,SU(3) \times SU(2) \times U(1)\,$,
~in order to obtain a spontaneous breaking 
of the global supersymmetry, and generate large tree-level masses 
for all squarks and sleptons. But the remaining unbroken continuous 
$\,R$-invariance discussed in the next section left us 
with massless gluinos, and the need,
later on, to generate a gluino mass either from radiative
corrections, or from supergravity.
In both cases the continuous $\,R$-invariance is reduced to its 
discrete $\,Z_2\,$ subgroup generated by the $\,R$-parity transformation, 
as we shall discuss in sections \ref{sec:rinv} and \ref{sec:grav}.},
acting as follows:

\vskip -.25truecm
\small
\be
\label{extra}
\!\!\left\{ \,
\ba{ccccl}
V(\,x,\,\theta,\,\bar\theta\ )&\rightarrow &  
V(\,x,\,\theta,\,\bar\theta\ )&&
\hbox{for the $SU(3) \times SU(2) \times U(1) $ gauge superfields;}  
\\ [.25truecm]
H_{1,2} (x,\,\theta) & \rightarrow  &
\ e^{-\,i\alpha}\ H_{1,2} (x,\,\theta)  &\ & 
\hbox{for the left-handed doublet Higgs superfields} \\
&&&& \hskip 2truecm H_1\ \hbox{and}\ H_2;  
\\  [.15truecm]
S (x,\,\theta) &\rightarrow  & e^{i\frac{\alpha}{2}}  \ 
S (x,\,\theta)  && 
\hbox{for the left-handed (anti)quark and (anti)lepton} \\ 
&&&& \hskip 2truecm  \hbox{superfields}\ \ \ Q, \,\bar U, \,\bar D,\ \,
L, \,\bar E.
\ea
\right.
\ee

\normalsize

\vskip .3truecm

\noindent
But a direct Higgs superfield mass term 
$\ \mu\ H_1 H_2\ $ in the superpotential
is {\it \,not\,} invariant under this extra $\,U(1)\,$ symmetry 
\,--\, nor is it under the continuous $\,U(1)$ $\,R$-invariance 
discussed in the previous and following sections.
Such a term, however, will get re-allowed, as soon 
as we shall abandon the extra $\,U(1)$ symmetry 
(given that no new neutral gauge boson or neutral-current interaction 
has been found),
and the continuous $\,U(1)\,$ $\,R$-invariance
(given that the gravitino and gluinos must be massive, as we shall discuss
in \hbox{section \ref{sec:grav}).}
\,The size of this 
``supersymmetric'' $\,\mu\,$ parameter may be naturally controlled
by using either the (broken) ``extra-$U(1)$'' symmetry (\ref{extra}),
or the continuous $\,R$-invariance, 
that must be broken at the same time as the supersymmetry.

\begin{table}[t]
\caption{\ Minimal particle content of the Supersymmetric Standard Model.
\label{tab:SSM}}
\vspace{0.2cm}
\begin{center}
\begin{tabular}{|c|c|c|} \hline 
&&\\ [-0.2true cm]
Spin 1       &Spin 1/2     &Spin 0 \\ [.1 true cm]\hline 
&&\\ [-0.2true cm]
gluons ~$g$        	 &gluinos ~$\tilde{g}$        &\\
photon ~$\gamma$          &photino ~$\tilde{\gamma}$   &\\ 
------------------&$- - - - - - - - - - $&--------------------------- \\
 

$\begin{array}{c}
W^\pm\\ [.1 true cm]Z \\ 
\\ \\
\end{array}$

&$\begin{array}{c}
\hbox {winos } \ \widetilde W_{1,2}^{\,\pm} \\ 
[0 true cm]
\,\hbox {zinos } \ \ \widetilde Z_{1,2} \\ 
\\ 
\hbox {higgsino } \ \tilde h^0 
\end{array}$

&$\left. \begin{array}{c}
H^\pm\\
[0 true cm] H\ \\
\\
h, \ A
\end{array}\ \right\} 
\begin{array}{c} \hbox {Higgs}\\ \hbox {bosons} \end{array}$  \\ &&\\ 
[-.1true cm]
\hline &&

\\ [-0.2cm]
&leptons ~$l$       &sleptons  ~$\tilde l$ \\
&quarks ~$q$       &squarks   ~$\tilde q$\\ [-0.3 cm]&&
\\ \hline
\end{tabular}
\end{center}
\end{table}

\vskip .3truecm
In any case, independently of the details of the
supersymmetry-breaking mechanism 
ultimately going to be considered, 
we obtain the following minimal particle content
of the Supersymmetric Standard Model, given in 
\hbox{Table \ref{tab:SSM}}.
Each spin-$\frac{1}{2}\,$ quark $\,q\,$ or charged lepton $\,l^-\,$
is associated with {\it \,two\,} spin-0 partners collectively denoted by
$\,\tilde q\,$ or $\,\tilde l^-\,$, ~while a left-handed neutrino $\,\nu_L\,$ 
is associated with a {\it \,single\,} spin-0 sneutrino $\,\tilde \nu$.
~We have ignored for simplicity 
further mixings between the various ``neutralinos''
described by neutral gaugino and higgsino fields, denoted in this Table
by $\,\tilde\gamma,\ \tilde Z_{1,2}$, 
and $\tilde h^0$.
More precisely, all such models include 4 neutral Majorana fermions at least,
corresponding to mixings of the fermionic partners of
the two neutral $\ SU(2) \times U(1)$ gauge bosons (usually denoted by 
$\,\tilde\gamma\,$ and $\,\tilde Z$, 
~or $\,\tilde{W_3}\,$ and $\,\tilde B\,$) ~and of the 
two neutral higgsino components 
($\,\tilde{h_1^{\,0}}\,$ and $\,\tilde{h_2^{\,0}}$). 
\,Non-minimal models also involve additional 
gauginos\footnote{If an extra $\,U(1)\,$
is gauged, one of the neutral Higgs bosons becomes an ``eaten'' 
Goldstone boson,
while the corresponding extra-$\,U(1)\,$ neutral gauge boson 
(called $\,Z'\,$ or $\,U\,$)
acquires a mass.} and/or higssinos.

\bigskip

\section{\sloppy
$R$-invariance and $\,R$-parity in the 
Supersymmetric Standard Model.}
\label{sec:rinv}

\vskip .1truecm

As we explained earlier, the early two-Higgs-doublet $\,SU(2) \times U(1)\,$ 
model of 1974~\cite{toy} \,showed how one could introduce 
a new $\,R\,$ quantum number, then defined as an 
{\it \,additive\,} quantum number (corresponding 
to a continuous $\,U(1)\,$ $\,R$-invariance)
carried by the supersymmetry generator,
and distinguishing between bosons and fermions 
inside the multiplets of supersymmetry. 
Gauge bosons and Higgs bosons
have $\,R=0\,$ while their partners under supersymmetry,
now to be interpreted as gauginos and higgsinos 
(rather than lepton field candidates), have $\,R=\,\pm1\,$.
The definition of this continuous $\,R$-invariance was then extended 
to the chiral quark and lepton superfields,
spin-$\frac{1}{2}\,$ quarks and leptons having $R=0$, 
~and their spin-0 superpartners, 
$\,R=+\,1\ $ (for $\,\tilde q_L,\,\tilde l_L\,$) ~or $\ R=-\,1\,$ 
(for $\,\tilde q_R,\,\tilde l_R\,$) ~\cite{ssm}.
~The action of these continuous $\,U(1)$
$R$-symmetry transformations, which survive the spontaneous breaking 
of the electroweak symmetry, is given in \hbox{Table \ref{tab:rinv}.}

\vskip .3truecm

\begin{table}[t]
\caption{\ Action of a continuous $\,U(1)\,$
$\,R$-symmetry transformation 
on the gauge and chiral superfields of the Supersymmetric Standard Model.
\label{tab:rinv}}
\vspace{0.2cm}
\begin{center}
\begin{tabular}{|cccl|}
\hline &&& \\
$\ V(\,x,\,\theta,\,\bar\theta\ )$&$\rightarrow$ &  
$V (\,x, \,\theta \,e^{-i\alpha},\,\bar\theta \, e^{i\alpha}\,)$ &
for the  $\,SU(3)\times SU(2)\times U(1)\,$ \\
&&&  \ \ \ \ gauge superfields   \\ [.3truecm]
$\ H_{1,2} \,(\,x,\,\theta\,)$ &$\rightarrow$  &
$H_{1,2}\, (\,x, \,\theta \,e^{-i\alpha}\,) $ & 
for the two left-handed chiral \\
&&&  \ \ \ \ doublet Higgs superfields \\  [.1truecm]
&&& \ \ \ \ $H_1\,$ and $\,H_2\,$  \\  [.3truecm]
$\ S (\,x,\,\theta\,)$ &$\rightarrow $ & $e^{i\alpha}  \ 
S (\,x, \,\theta \,e^{-i\alpha}\,)$  & for the left-handed chiral \\
&&&  \ \ \ \ (anti)quark and lepton  \\  [.1truecm]
&&&   \ \ \ \ superfields $\,Q, \,\bar U, \,\bar D,\ \,
L, \,\bar E $ \\  [.3truecm]
\hline
\end{tabular}
\end{center}
\end{table}

The $\,SU(3)\times SU(2)\times U(1)\,$ gauge interactions of the chiral 
quark and lepton superfields, and of the two doublet Higgs 
superfields $\,H_1\,$ and $\,H_2\,$,
are indeed invariant under this continous $\,U(1)\,$ $\,R$-symmetry. 
\,So are the super-Yukawa interactions of the
two doublet Higgs superfields $\,H_1$ and $\,H_2$
responsible for the generation of quark and lepton masses
through the superpotential (\ref{supot}).
\,Indeed this trilinear superpotential 
transforms under continuous 
$\,R$-symmetry with ``$\,R$-weight'' $\ n_{\cal W}\,=\,\sum_i\,n_i\,=\,2\,$,
~i.e. according to
\be
{\cal W}\,(\,x,\,\theta\,) \ \ \rightarrow\ \  e^{2\,i\alpha}\ \ 
{\cal W}\,(\,x, \,\theta \,e^{-i\alpha}\,)\ \ .
\ee

\noindent
Its auxiliary ``$\,F$-component'' is therefore $\,R$-invariant, 
and generates $\,R$-inva\-riant interaction terms 
in the Lagrangian density.

\vskip .3truecm

This $\,R$-invariance led us to distinguish betwen a sector of 
$\,R$-{\it even particles\,},
which includes all the ordinary particles of the standard model, 
gauge and Higgs bosons, leptons and quarks,
with $\,R\,=\,0\,$;
~and their $\,R$-{\it odd superpartners\,}, gauginos and higgsinos, 
and spin-0 leptons and quarks, with $\,R\,=\,\pm \,1\,$, 
as indicated in Table \ref{tab:Rp}. 

\vskip .3truecm
More precisely the necessity of generating masses for the (Majorana)
\hbox{spin-$\frac{3}{2}\,$} gravitino~\cite{ssm2}, 
and for the spin-$\frac{1}{2}\,$ gluinos,
as we shall discuss later in section \ref{sec:grav},
did not allow us to keep the distinction between $\,R\,=\,+\,1\,$
and $\,R\,=\,-\,1\,$ particles, forcing us to abandon the continuous 
$\,R$-invariance in favor of its discrete version, $\,R$-parity.
~The \,-- even or odd --\, parity character of 
the (additive) $\,R\,$ quantum number 
corresponds to the well-known $\,{\boldmath R}${\bf \,-parity}, 
first defined as $\,+\,1\,$ for the ordinary particles
and $\,-\,1\,$ for their superpartners, 
\,which may be written as $\,(\,-\,1\,)^{R}\ $~\cite{rp}:
\be
\label{rp1}
\hbox {\framebox [11cm]{\rule[-.5cm]{0cm}{1.2cm} $ \displaystyle {
R\hbox{-parity}\ \ \,R_p\ \ =\ \ (\,-\,1\,)^{R}\ \ =\ \ \left\{ \ 
\ba{l} 
+\,1\ \ \ \ \hbox{for ordinary particles,} \vspace{2mm} \\
-\,1\ \ \ \hbox{for their superpartners.}
\ea  \right.
}$}}
\ee

\begin{table}[t]
\caption{\ $R$-parities 
in the Supersymmetric Standard Model.
\label{tab:Rp}}
\vspace{0.2cm}
\begin{center}
\begin{tabular}{|c|c|} \hline 
&\\ [-0.2true cm]
Bosons       & Fermions     \\ [.2 true cm]\hline \hline 
&\\ [-0.1true cm]
\ gauge and Higgs bosons\ \ (\,$R\,=\,0$\,) \ & 
\ gauginos and higgsinos\ \ (\,$R\,=\,\pm\,1$\,)  \ \\ & \\[-0.2cm]
$R$-parity\ \ +  & $R$-parity\ \ {$-$} \\ & \\ \hline & \\[-0.1cm]
sleptons and squarks\ (\,$R\,=\,\pm\,1$\,) & 
leptons and quarks\ \ (\,$R\,=\,0$\,)  \\ [-0.2 cm] & \\
$R$-parity\ \ $-$  & $R$-parity\ \ $+$ \\ & \\ \hline
\end{tabular}
\end{center}
\end{table}

\bigskip

\smallskip

\section{Relation of $\,R$-parity with $\,B\,$ and $\,L\,$ quantum numbers.}
\label{sec:rbl}

\vskip .1truecm
In addition, there is a close connection between $\,R$-parity 
and baryon and lepton number conservation laws, 
which has its origin in our initial desire 
to obtain supersymmetric theories in which $\,B\,$ and $\,L\,$ 
could be conserved. 
Actually the superpotential of the theories discussed in Ref. \cite{ssm}
\,was constrained from the beginning, for that purpose,
to be an {\it \,even\,} function of the quark and lepton superfields.

\vskip .3truecm
In other terms, {\it\,odd\,} superpotential terms 
\,($\,{\cal W}\,'$),
~which would have violated the ``matter-parity'' symmetry 
$\,(\,-1)^{(3B+L)}$, ~were excluded from the beginning,
to be able to recover $\,B\,$ and $\,L\,$ conservation laws,
and at the same time 
avoid unwanted direct Yukawa exchanges of spin-0 quarks and leptons
between ordinary spin-$\frac{1}{2}\,$ quarks and leptons.
Tolerating unnecessary superpotential terms which are {\it odd\ } 
functions of the quark and lepton superfields
\,(i.e. $R_p$-violating terms, such as those which 
were widely discussed later), 
does indeed create immediate problems
with baryon and lepton number conservation laws:
most notably a squark-induced proton instability 
with a much too fast decay rate, if both 
$\,B\,$ and $\,L\,$ violations are simultaneously allowed; or neutrino 
masses (and other effects) that could be too large, 
if $\,L$-violations are allowed 
so that ordinary neutrinos can mix with neutral higgsinos and gauginos.

\vskip .3truecm
The question was raised very early, in 
the discussion of the phenomenology of supersymmetric theories, 
and the experimental searches for the gluinos and the \hbox{$``R$-hadrons''}
they could form~\cite{ff,ff2},
of how general is this notion of \hbox{$R$-parity,} as defined previously
by eq.\,(\ref{rp1}).
To answer more easily it is useful to make the above connection between
$\,R$-parity and  $\,B\,$ and $\,L\,$ conservation laws more transparent.
It can indeed be made quite obvious by noting that
for usual particles, $\ (-1)\,^{2 \ \hbox{\footnotesize Spin}}$ coincides 
with $\ (-1)\,^{3B+L}\,$.
~This immediately leads to a simple redefinition of the $\,R$-parity (\ref{rp1})
in terms of the spin $\,S\,$ and a ``matter-parity'' $(-1)\,^{3B+L}\,$, 
~as follows~\cite{ff}:
\be
\label{rp2}
\hbox {\framebox [6.5cm]{\rule[-.5cm]{0cm}{1.2cm} $ \displaystyle {
R\hbox{-parity} \ \ =\ \ (-1)\,^{2S} \ (-1)\,^{3B+L}    \ \ .     
}$}}\ee
This may also be rewritten as $\ (-1)^{2S} \ (-1)\,^{3\,(B-L)}\,$, 
~showing that $R$-parity may still be conserved 
even if baryon and lepton numbers 
are separately violated (as in grand-unified theories), 
as long as their difference ($\,B-L\,$) remains 
conserved, even only modulo 2.

\vskip .3truecm
This $\,R$-parity symmetry operator
may also be viewed as a non-trivial geometrical discrete symmetry 
associated with 
a reflection of the anticommuting fermionic Grassmann coordinate, 
$\,\theta\ \to -\,\theta\,$, in superspace~\cite{geom}.

\vskip .3truecm

This $\,R$-parity operator plays an essential r\^ole in the construction 
of supersymmetric theories of interactions, 
and in the discussion of the experimental signatures of the new particles.
$\,R$-invariance or simply its 
discrete version, a conserved $\,R$-parity, 
guarantees that {\it \,the new spin-0 squarks and sleptons 
cannot be directly exchanged\ } between ordinary quarks and leptons.
But let us now discuss more precisely the reasons
which led us to abandon the continuous $\,R$-invariance 
in favor of its discrete version, $\,R$-parity.

\medskip

\vskip 2truecm

\section{Gravitino and gluino masses: from $\,R$-invariance to $\,R$-parity.}
\label{sec:grav}

\vskip .1truecm
There are at least two strong reasons to abandon, at some point, 
the continuous $\,R$-invariance, in favor of its discrete $\,Z_2\,$ 
subgroup generated by the $\,R$-parity 
transformation\footnote{We disregard  
other possibilities involving extended supersymmetry
with a continous $\,U(1)$ $\,R$-invariance, 
which may allow for massive {\it \,Dirac\,} gravitinos and gluinos, 
carrying one unit of a conserved, additive $\,R\,$ quantum number.
}.
One is theoretical, the necessity \,-- once gravitation is introduced --\,
of generating a mass for the 
(Majorana) spin-$\frac{3}{2}\,$ {\it \,gravitino\,\,} in the 
framework of sponta\-neously-broken locally supersymmetric theories~\cite{ssm2}.
The other is phenomenological, the 
non-observation of massless (or even light) {\it \,gluinos}.
Both particles would have to stay massless, in the absence of 
a breaking of the continuous $\,U(1)\,$ $\,R$-invariance.

\vskip .3truecm
This is also connected with the mechanism 
by which the supersymmetry should get spontaneously broken, 
in the Supersymmetric Standard Model.
The question still has not received a definitive answer yet.
While we first considered in 1976 the inclusion of
universal soft supersymmetry-breaking terms for all squarks and sleptons,
\be
-\ \sum_{\tilde q,\,\tilde l}\ m_0^{\,2}\ \ 
(\,{\tilde q}^\dagger\,\tilde q\ +\ {\tilde l}^\dagger \,\tilde l\,)\ \ ,
\ee
such terms should in fact be generated by some spontaneous 
supersymmetry-breaking mechanism, if supersymmetry is to be realized locally.

\vskip .3truecm
Indeed they could be generated spontaneously, for example
by gauging the ``extra-$U(1)\,$'' symmetry (\ref{extra})
already mentioned in \hbox{section \ref{sec:ssm}.}
\,This symmetry is associated, in the simplest case, 
with a purely axial extra-$U(1)$ current 
for all quarks and charged leptons.
Gauging such an extra $\,U(1)\,$ is in fact 
necessary~\cite{extrau},
\,if one intends to generate large positive mass$^2$ 
for all squarks ($\,\tilde u_L,\,\tilde u_R,\ \tilde d_L,\,\tilde d_R\,$)
and sleptons, at the classical level, in a spontaneously-broken
globally supersymmetric theory.
But this required new neutral current interactions \,-- unobserved --\,
and left us with the necessity of generating, also, large gluino masses 
-- a question to which we shall return soon.
As a result, the gauging of an extra $U(1)\,$ no longer appears
as an appropriate way to generate large superpartner masses.
One now uses in general, again, soft supersymmetry-breaking 
terms~\cite{gg} (possibly ``induced by supergravity''),
which essentially serve as a parametrization 
of our ignorance about the 
true mechanism of supersymmetry breaking chosen by Nature
to make superpartners heavy (if supersymmetry is indeed 
a symmetry of Nature\,!).

\vskip .3truecm
Let us return to the question of gluino masses.
Since $\,R$-transformations act {\it \,chirally\,} 
on the Majorana octet of gluinos,
\be
\tilde g\ \ \to\ \ e^{\,\gamma_5\,\alpha}\ \tilde g\ \ .
\ee
a continuous $\,R$-invariance
would require the gluinos to remain massless,
even after a spontaneous breaking of the supersymmetry\,!
We would then expect the existence of 
relatively light $\,``R$-hadrons'' 
(bound states
of quarks, antiquarks and gluinos), 
which have not been observed~\cite{ff,ff2}.
Present experimental results indicate that gluinos, if they do exist, 
must be very massive, requiring a significant 
breaking of the continuous $\,R$-invariance.

\vskip .3truecm
In the framework of global supersymmetry
it is not so easy to generate large gluino masses.
Even if global supersymmetry is spontaneously broken, 
and if the continuous $R$-symmetry is not present, 
it is still in general rather difficult to obtain large masses for gluinos, 
since~\cite{glu}: 

\vskip .1truecm
i) no direct gluino mass term is present in the 
supersymmetric Lagrangian density; 

\vskip .1truecm
ii) no gluino mass term may be generated spontaneously, at the tree 
approximation: gluino couplings involve {\it colored} spin-0 fields, 
which cannot be translated if the color $\,SU(3)\,$ gauge group
is to remain unbroken;

\vskip .1truecm
iii) a gluino mass term may then be generated by radiative corrections,
but this can only be through diagrams which ``know'' both about:

\hskip .5truecm
a) the spontaneous breaking of the global supersymmetry,
through some appropriately-generated 
$\,<\!D\!>,\ <\!F\!>\,$ or $\,<\!G\!>\,$,
\,as discussed in \hbox{section \ref{sec:na}};

\hskip .5truecm
b) the existence of superpotential interactions which do not preserve 
the continuous $\,U(1)\,$ $\,R$-symmetry.

\vskip .3truecm
Ref. \cite{glu} showed that it was indeed possible to generate 
gluino masses by radiative corrections, 
through the interaction of gluinos with an ``ad hoc'' 
sector of what would be called now {\it \,vectorlike ``messenger'' quarks\,}, 
sensitive to the spontaneous breaking of the supersymmetry.
But gluino masses radiatively generated along these lines 
generally tend to be rather small, 
unless one accepts to introduce, in some (often rather complicated)
``hidden sector'', very large mass scales $\,\gg\,m_W$, 
~so that radiatively-generated gluino masses could still end 
up to be of the order of several hundreds of GeV/$c^2$ 's, 
~as now experimentally required.

\vskip .5truecm

     Fortunately gluino masses may also result directly from 
supergravity, as already observed in 1977~\cite{ssm2}. Gravitational 
interactions require, within local supersymmetry, that the spin-2 
graviton be associated with a spin-$3/2\,$ partner~\cite{sugra}, the 
gravitino. Since the gravitino is the fermionic gauge particle of 
supersymmetry it must acquire a mass, $\,m_{3/2} \ (= \kappa \ d/ \sqrt{6}
 \ \approx \,
d/m_{\rm{Planck}}\,\,)$, as soon as the local supersymmetry gets spontaneously 
broken. 
Since the gravitino is a self-conjugate Majorana fermion
its mass breaks the continuous $\,R$-invariance which acts chirally on it
(just as for the gluinos)~\cite{ssm2}, 
\,forcing us to abandon the continuous $U(1)$ $R$-invariance, 
in favor of its discrete $\,Z_2\,$ subgroup 
generated by the $R$-parity transformation.
We can no longer distinguish between the values +1 and $-1$ of the
 ({\it additive}) quantum number $R$; 
\,but only between \hbox{``$R$-odd''} particles 
(having $\,R\,=\,\pm1\,$) ~and ``$R$-even'' ones, ~i.e. between 
particles having $R$-parities $\,R_p\,=\,(-1)^R\,=\,-\,1,$ ~and $\,+\,1$, 
\,respectively, 
as indicated in section \ref{sec:rinv} (cf. \hbox{Table \ref{tab:Rp}}).

\vskip .6truecm
In particular, when the spin-$\frac{3}{2}\,$ gravitino mass term 
$\,m_{3/2}\,$ is introduced, the ``left-handed sfermions''
$\,\tilde f_L$, ~which carry $\,R\,=\,+\,1$, ~can mix with the 
right-handed'' ones  $\,\tilde f_R$, 
~which carry $\,R\,=\,-\,1$, ~through mixing terms having 
$\ \Delta \,R\,=\,\pm \,2\,$, ~which may naturally \,(but not necessarily)
be of order $\ m_{3/2}\ m_f\,$.
Supergravity theories offer a natural framework 
in which to include direct gaugino Majorana mass terms 
\be
-\ \frac{i}{2}\ \ m_3\ \ \bar {\tilde G}_a\,\tilde G_a\ \
-\ \frac{i}{2}\ \ m_2\ \ \bar {\tilde W}_a\,\tilde W_a\ \
-\ \frac{i}{2}\ \ m_1\ \ \bar {\tilde B}\,\tilde B\ \ ,
\ee
which also correspond to $\,\Delta \,R\,=\,\pm\, 2\,$.
~The mass parameters $m_3,\ m_2\,$ and $\,m_1$, 
~for the $\,SU(3) \times SU(2) \times U(1)\,$ gauginos,
could naturally \,(but not necessarily)\,
be of the same order as the gravitino mass $\,m_{3/2}\,$.
\,This directly leads us to \hbox{$R$-parity,} defined as $\,R_p\,=\,(-1)^R$, 
~as indicated in section \ref{sec:rinv},
$R$-parity being +1 for ordinary particles, and $-1$ 
for their superpartners.
Of course, once the continuous $R$-invariance 
is reduced to its discrete $\,R$-parity subgroup, 
a direct Higgs superfield mass term
$\ \,\mu \ H_1 H_2\ $ may be re-allowed  
in the superpotential, as done for example in the MSSM.

\vskip .6truecm

     In general, irrespective of the supersymmetry-breaking mechanism
considered, one normally expects superpartners not to be too heavy. 
Otherwise the corresponding new mass scale would tend to 
contaminate the electroweak scale, thereby
{\it creating\,} a hierarchy problem in the Supersymmetric Standard Model.
Superpartner masses are then normally expected to be naturally of the
order of $\,m_W$, ~or at most in the $\ \sim\,$ TeV$/c^2\,$ range.

\section{Conclusions.}

\vskip .1truecm

The Supersymmetric Standard Model (``minimal'' or not),
with its $R$-parity symmetry (absolutely conserved, or not),
provided the basis for the experimental searches for the new superpartners
and Higgs bosons. 
However, the ``final'' answer about how the supersymmetry 
should actually be broken is not known, 
and this concentrates most of the remaining uncertainties in the
Supersymmetric Standard Model.
Since the first searches for gluinos and photinos, 
selectrons and smuons, starting in the years 1978-1980, 
the experimental efforts have been pursued incessantly, 
for more than twenty years now, 
without giving us any direct evidence for the new supersymmetric parti\-cles 
yet. Supersymmetry as a symmetry of the real world, 
and the existence of the superpartners, 
and of the new additional Higgs bosons, still remain as 
physical hypotheses, that we would like to see confirmed experimentally, 
some day.

\vskip .3truecm

Many good reasons to work on supersymmetry, the 
Supersymmetric Standard Model and its various possible extensions 
have been widely discussed, 
dealing with supergravity, grand-unification
(the $\,SU(3) \times SU(2) \times U(1)\,$ gauge couplings 
tend to unify at high energies, 
when their evolution is computed with the field content 
of the Supersymmetric Standard Model),
\,extended supersymmetry, new spacetime dimensions,
superstrings, \hbox{``$M$-theory'',} ... .
\,But, after more than 20 years of experimental searches, 
we would certainly appreciate 
to start seeing the missing half of the SuperWorld 
being disclosed experimentally\,!

\end{document}